\documentclass[manuscript,utf8]{acmart}
\usepackage{CJKutf8}

\AtBeginDocument{%
  \providecommand\BibTeX{{%
    \normalfont B\kern-0.5em{\scshape i\kern-0.25em b}\kern-0.8em\TeX}}}

\setcopyright{acmcopyright}
\copyrightyear{2024}
\acmYear{2024}
\setcopyright{rightsretained}
\acmConference[DIS '24]{Designing Interactive Systems Conference}{July 1--5,
2024}{IT University of Copenhagen, Denmark}
\acmBooktitle{Designing Interactive Systems Conference (DIS '24), July 1--5,
2024, IT University of Copenhagen,
Denmark}\acmDOI{10.1145/3643834.3660690}

\acmConference[Conference acronym 'XX]{Make sure to enter the correct
  conference title from your rights confirmation emai}{June 03--05,
  2018}{Woodstock, NY}
\acmPrice{15.00}
\acmISBN{978-1-4503-XXXX-X/18/06}

\begin{document}

\title{The Power of Absence: Thinking with Archival Theory in Algorithmic Design}



\author{Jihan Sherman}
\email{jihan@estelleandboots.com}
\orcid{https://orcid.org/0000-0002-1578-566X}
\affiliation{%
  \institution{Estelle + Boots}
  \streetaddress{}
  \city{Atlanta}
  \state{GA}
  \country{United States}}

\author{Romi Morisson}
\email{rmorrison28@gmail.com}
\orcid{https://orcid.org/0009-0002-1415-2374}
\affiliation{%
  \institution{Digital Media Arts, University of California, Los Angeles}
  \streetaddress{}
  \city{Los Angeles}
  \state{CA}
  \country{United States}}

\author{Lauren Klein}
\email{lauren.klein@emory.edu}
\orcid{https://orcid.org/0000-0003-2603-878X}
\affiliation{%
  \institution{Quantitative Theory \& Methods, Emory University}
  \streetaddress{}
  \city{Atlanta}
  \state{GA}
  \country{United States}}

\author{Daniela K. Rosner}
\email{dkrosner@uw.edu}
\orcid{https://orcid.org/0000-0001-9448-914X}
\affiliation{%
  \institution{Human Centered Design \& Engineering, University of Washington}
  \streetaddress{}
  \city{Seattle}
  \state{WA}
  \country{United States}}

\renewcommand{\shortauthors}{Anonymous}

\begin{abstract}
This paper explores the value of archival theory as a means of grappling with bias in algorithmic design. Rather than seek to mitigate biases perpetuated by datasets and algorithmic systems, archival theory offers a reframing of bias itself. Drawing on a range of archival theory from the fields of history, literary and cultural studies, Black studies, and feminist STS, we propose absence—as power, presence, and productive—as a concept that might more securely anchor investigations into the causes of algorithmic bias, and that can prompt more capacious, creative, and joyful future work. This essay, in turn, can intervene into the technical as well as the social, historical, and political structures that serve as sources of bias.
\end{abstract}




\maketitle

\section{Introduction}
Nearly a decade into the Fourth Industrial Revolution—an era characterized by the convergence of several emerging technologies such as cloud computing, blockchain, generative AI, and other algorithmic systems—the issue of algorithmic bias has entered mainstream conversation \cite{news-02,news-03,fuller_algorithm_2008}. Algorithmic bias generally refers to the uneven effects of algorithmic systems driving the automation and digitization of industries, often reflecting and reproducing social hierarchies marked by race, class, disability, sexuality, gender and other categories of difference. These systems include facial recognition algorithms that have higher error rates for individuals with darker skin tones \cite{nkonde2019automated}; online advertising platforms that perpetuate discriminatory practices in housing and job advertisements \cite{noble_algorithms_2018}; credit scoring algorithms that reanimate the exclusion of racialized groups from accessing credit, loans, and other financial services \cite{eubanks_automating_2018}; and predictive healthcare technologies that result in misdiagnosis, inadequate treatment, and uneven access to care \cite{johnson2023accuracy}. 

This range of concerns grows in large part from advocacy efforts of groups like the Algorithmic Justice League \cite{alg-justice}, investigative reporting by organizations like ProPublica \cite{propub-01,propub-02}, academic research by scholars such as Safiya Noble \cite{noble_algorithms_2018}, Ruha Benjamin \cite{benjamin_machine_2021,benjamin_race_2019}, and Meredith Broussard \cite{broussard2018artificial}, among others, and newly enacted regulation by the European Union \cite{euro-bias,lampinen_cscw_2022} and more recently, the White House \cite{news-01}. Even the IEEE Global Initiative on Ethics of Autonomous and Intelligent Systems has weighed in, developing a set of design guidelines that emphasize the need for transparency, accountability, and a focus on human well-being \cite{ieee-bias}. Along with this work, which has highlighted the individual and social harms of discriminatory algorithms if left unchecked, has come a range of proposals from within the field of computer science for mitigating their biases and their pernicious effects. These include proposals for documenting the composition of datasets (e.g. ``Datasheets for Datasets'' \cite{gebru2022excerpt}) and the performance of models (e.g. ``Model Cards for Model Reporting'' \cite{mitchell2019model}), heightened standards for data sharing and research replicability \cite{rahrooh2024towards,kapoor2023leakage}, and mechanisms for enhanced interpretability and explainability of the algorithms themselves (e.g. ``Honest Students'' \cite{eisenstein2022honest}). 

Each of these projects represent admirable–and necessary–interventions into the issue of algorithmic bias. Their practical proposals complement work from the fields of design research, critical data studies, STS, and HCI, among others, which have proposed a range of analytical methods for examining, challenging, and attempting to change the unequal structural power at the root of algorithmic bias \cite{morrison2022voluptuous,benabdallah2022slanted,rosner2022bias}.

But more work is required if we are to be able to intervene into the issue of bias at its source. As these scholars inform us, the source of algorithmic bias is not any particular dataset or model, but instead the structural power differentials that produce social inequality—legacies of capital and colonial power already baked into practices of employment and hiring, healthcare, credit scoring, advertising and more. It is this social inequality that, in turn, produces biased datasets; and that prohibits any single approach to algorithmic ``fairness'' from achieving its intended effect. 

Looking closer at algorithmic inequities in the access of loans, receipt of medical treatment, or estimated risk of recidivism, we find that bias cannot be dismissed or corrected away. The issue lies in the relationship to historical data—to lives and worlds left undocumented, unsaid, and untold. A criminal justice or healthcare system may have never existed without the inequities of the present, and yet such a possibility remains. We have no documentation, no traditional evidence or material to validate that existence; but we do have the capacity to imagine that omission—that absence—otherwise. Rather than theorizing algorithmic systems as problems to solve, we bring a reframing of the issue of algorithmic bias itself. We propose absence as a concept that might more securely anchor investigations into the causes of algorithmic bias as well as prompt more generative, more capacious, and more creative interventions into the technical as well as the social, historical, and political structures that serve as the sources of bias. 

This paper describes the idea of absence as it has been theorized through the field of archival studies. Archives, after all, are early instantiations of datasets. Like the collections of web pages and images used to train today’s generative AI systems, the contents of archives also consist of documents, images, and other artifacts, compiled and preserved because of what they can tell us about cultures past and present. As with today’s datasets, archives are also sometimes very intentionally curated, as is the case with the archive of the Papers of Thomas Jefferson, which was initially assembled with the help of a custom copy-machine that Jefferson himself designed, and used to select the documents that would be individually copied and preserved; or they are created with whatever contents can be found, as with the Internet Archive, which combines web-crawling with other collection methods including user uploads and organizational partnerships. Regardless of the degree of curation, no archive—or dataset—is ever fully and perfectly complete. We see this incompleteness with clarity in artist Mimi Onuoha's \textit{Library of Missing Datasets} wherein the artist creates a physical file cabinet as a repository of information that has been overlooked ``in a society where so much is collected'' \cite{missing-datasets}. And herein lies the contribution of the idea of absence for technical researchers: it provides a way of thinking through the biases in the data that cannot ever be fixed, and in so doing, provides a path for moving forward with technical work that allows the data and its biases to both be taken into account.

In this paper we summarize how the design research community has engaged with bias, both practically and theoretically, and provide a genealogy of the idea of absence in archival research.  Suspending the dictates of linearity, we tend to the past to provide a series of design speculations which illustrate how absence takes shape across two sites, and how archival theory enabled the scholars who encountered those absences in their work to move beyond them. We close with a reflection on bodily knowing and the possibility of building that sensorial perception into algorithmic design.

\section{Methods}
Our methods are rooted in a range of critical humanist traditions, including feminist methodologies of situated inquiry \cite{d2023data}, programs of design speculation that interrogate alternate futures \cite{rosner2018critical}, and generative archival analyses grounded in Black studies \cite{sherman_black_2023,morrison_gaps_2019}. Contrasting with a teleological, progressive analysis of historical material, we draw from a genealogical approach that foregrounds alternative pathways of interpretation and theorizing. Where teleological analysis tends to view history as a coherent and linear progression, genealogical analysis focuses on the complex and often discontinuous development of ideas, practices, and performances \cite{denton_bringing_2020}. Genealogies involve the critical examination of power, coloniality, and knowledge that have shaped social and historical phenomena, with particular attention to suppressed narratives that complicate hegemonic historical accounts \cite{denton2021genealogy,valdivia2023datafication}. Building from the historical analysis of Michel Foucault \cite{foucault2012archaeology}, and reworked through the alternative genealogies of Sylvia Wynter \cite{wynteropen}, Hortense Spillers \cite{spillers2022mama}, Saidiya Hartman \cite{hartman2008venus}, among others, this approach creates the groundwork for examining what Alexander Weheliye has called the ``ideological and physiological mechanics of the violently tiered categorization of the human species in western modernity'' (\cite{weheliye2014habeas}, p.29).

Our analysis draws from these interconnected perspectives to offer a detailed reframing of algorithmic bias along three lines of argumentation: one focused on absence as power, a second on absence as presence, and a third on absence as productive. With this three-part analysis, we probe two central questions. First, we ask: How does bias change as a result of working with absence? Next, we ask: What does absence open up or foreclose for design?

\section{Background}

\subsection{Archives and Algorithms}
As we turn toward theoretical developments in archival theory with relevance to design research, we look to the growing body of work that has turned to archival practices and related practices of documentation and curation as ways to reckon with the biases of algorithmic systems and data practices. 

The strand of this work that most directly connects to this paper has focused on practical implications of treating AI/ML datasets as archives, including changes to archiving practices \cite{colavizza2021archives}. Sometimes gathered under the umbrella of Computational Archival Science \cite{marciano2018archival,lemieux2019blockchain}, this range of work has argued for responding to digital developments with updates to archival thinking, such as paying greater attention to trust, usability, and archival context \cite{ranade2018access,colavizza2021archives}. In Giovanni Colavizza and colleagues’ survey of AI-supported recordkeeping research, they introduce a number of considerations, including the automation of record-keeping processes, the handling of sensitive information, and the growing capacity for archives to accommodate novelty and scale. Eun Seo Jo and Timnit Gebru’s landmark paper outlines five central document collection approaches—consent, power, inclusivity, transparency, and ethics and privacy—that have long term implications for data collection and annotation methodologies within algorithmic systems research, ultimately proposing that computer scientists learn best practices from archivists themselves \cite{jo2020lessons}. Thylstrup and colleagues \cite{thylstrup2021uncertain} compile a glossary of essay-length data ruminations that bring critical archival theory into conversation with the limits of datafication.

A second strand of related work focuses on providing genealogies of datasets in ways that explicitly and implicitly make use of archival practices. A growing group of scholars have focused on understanding how prominent datasets come to be  \cite{mei2023bias,denton_bringing_2020}. In their analysis of ImageNET, for example, Emily Denton and colleagues find that assumptions around ImageNet and other large computer vision datasets more generally rely on data aggregation and accumulation, the computational construction of meaning, and the process of making certain types of data labor invisible \cite{denton2021genealogy}. 
Ana Valdivia and Martina Tazzioli call this process of making invisible one of the ``racialising classifications and exclusionary mechanisms enforced through datafication'' (the racializing effects of quantification) \cite{valdivia2023datafication}. 

A final strand of work engages the effects of ``quality'' filtering. Jesse Dodge et al. \cite{dodge2021documenting,elazar2023s,zhou_rating_2009} have undertaken a related set of projects to document the large text-based C4 (Colossal Clean Crawled Corpus \cite{raffel2020exploring}) dataset that is used to train many LLMs. In addition to tracing as many of the documents it contains directly to their sources \cite{arnett2006gee}, they have also revealed the effects of blocklist filtering, which disproportionately removes text about minoritized people \cite{dodge2021documenting}. Related work has shown how quality filtering also likely impacts contents from lower SES and rural communities \cite{gururangan2022whose}. These projects show the value of engaging in such media-archaeological work, and suggest how a richer theorization of absence could help researchers move forward in the wake of such findings. 

Our work grows from these connected interests in the archive but from a theoretical perspective. While overwhelmingly focused on practice, this existing work has sometimes engaged the archive through theoretical precepts like ``provenance'' \cite{kale2023provenance} and ``original order'' \cite{colavizza2021archives}, concepts that emphasize the practical conditions and methods that define archival units. Rather than trace archival lessons for AI/ML design, or surface digital repercussions for archival practice, our work investigates algorithmic systems vis-a-vis archival theory. We focus on the underlying conceptual tools within critical archival theory that reshape methodological developments, including our design engagements with bias.

\subsection{Design and Bias}
While the topic of bias has been present within design research, this scholarship has largely emphasized technical procedures such as uneven forms of data collection \cite{hellman2021big,waldman2019power} and discriminatory annotation practices \cite{okidegbe2021discredited}. In this analysis, algorithmic bias has tended to sit within a defined set of conditions whose problems might be amended, fixed, or otherwise disbanded if constructed otherwise. One emphasis of this work has been on the construction and assembly of datasets, both in terms of who is enrolled to annotate and organize them and how aspects of the world get turned into data. This work analysis has involved problematizing not only the selection process (what counts as data) but also the influence of the recording studio environment, including how people feel when their spoken words get turned into training data \cite{mengesha2021don}. Other work challenges the solutionist impulse of fixing the technology development process (including more or different data, changing how recordings get circulated, or who directs the process) \cite{ngueajio2022hey,cunningham2023grounds}. This work tends to point to regulatory structures that might better address the fact that harmful bias and epistemic violence attributed to datasets or algorithmic systems is already embedded in---and reflects back---wider structural inequities within society \cite{benjamin_machine_2021,noble_algorithms_2018}.  

An important area of design research closely connected to our own offers a complementary response to the entrenched inequity influencing and amplified by algorithmic developments. Drawing from speculative approaches, this work has sought to use artistic and exploratory experimentation with algorithmic systems and related data practices to probe the limits and potentialities of ``bias,'' material and cultural phenomenon. In a series of works \cite{rosner2023bias,rosner2022bias,benabdallah2022slanted,lustig_tarot_2022}, Gabrielle Benabdallah, Chari Glogovac-Smith, Afroditi Psarra, Caitlin Lustig and several others have examined ``bias'' through textile etymologies, where the term refers to the angular cross grain of woven fabric. With exercises such as ``cutting on the bias'' \cite{rosner2022bias} (the diagonal) of fabric to create a data set, they examine how the production of data requires making choices about where to make cuts or lines between what is in and out---reflecting both the environment and the dataset creator (their assumptions, skills, tools etc). Our analysis directly builds from this analysis of material bias in our engagement with design speculations informed by critical archival theory and the genealogies of absence they help conjure.

\section{A Genealogy of Absence}
Our genealogy of absence derives from the fields of literary and archival theory, as elaborated through feminist theory and Black studies. In this work we find a rich history of scholars and practitioners who intentionally attune themselves to the absences in the archive, exploring the reasons for those absences as well as the various methods that might be employed in order to invest those absences with meaning. 

\subsection{What is Absence?}
In common parlance, absence refers to a lack or non-existence. This paper departs from this narrow conception of absence to consider a wider constellation of meanings. We are particularly informed by a range of work at the nexus of critical archival theory and Black feminist thought to position absence as an active presence, an influence despite its non-physical manifestation. This definition implies that even though something (data, algorithmic processes, documentation) may not be physically present, its impact or influence is still felt and recognized. By shifting our focus from bias to absence, we gain a more capacious concept with which to anchor questions of algorithmic harm. More concretely, this shift enables us to see the practical limits of approaches to bias that frame their interventions as ``fixes'' for the flaws of such datasets and data systems. Such fixes, which often focus on shifting visible outputs, have the effect of naturalizing the relationship between bias and corrective, rather than retaining focus on the source of the bias in the first place. Reconceptualizing bias as absence, however, enables us as researchers to train our attention on the reasons for the bias to enter into the dataset or data system. It points us to the social, historical, and political conditions that give rise to biased data, as well as to the spaces of indeterminacy – and to the irrecoverably missing parts – of any dataset or system. It urges analysts away from viewing the dataset as a binary state (biased or non-biased) and toward viewing all datasets as embedded in specific relationships of power. This recognition of power follows Yanni Loukissas’ important differentiation between the dataset and data settings, which draws attention to the material, historical, and social contexts from which information is cleaned, organized, and relegated into interoperable units. In addition, framing bias as absence opens up questions of both social responsibility and personal commitment: what responsibility does each of us have as a researcher working with a particular dataset, and what are our commitments to pursuing knowledge in that space?   


\subsection{Absence as Power}
Here we might start with the work of Haitian American anthropologist Michel-Rolph Trouillot: 

\begin{quote}
   ``The moment of fact creation (the making of sources); the moment of fact assembly (the making of archives); the moment of fact retrieval (the making of narratives); and the moment of retrospective significance (the making of history) in the final instance''  \cite{trouillot2015silencing}.
\end{quote}

 While Trouillot presents these moments as discrete, he later explains how each is returned to again and again as each scholar attempts to make meaning from the documents that the archive contains (or does not contain, as the case may be). Trouillot takes as his focus the Haitian Revolution, so the process of ``the making of sources'' is bound up in colonial power relations between French citizens and Haitian subjects, which are of course also racial power relations because of the dominant identities of each group. Colonial governments produce documents about colonial subjects; they have the institutional, education, and physical apparatus that enables them to create records of their lives, while colonial subjects largely do not. This moment in which power enters and shapes the contents of the archive takes place even before the archive is created. By the same token, we might recognize how datasets are shaped by unequal power even before they are recognized as such. 

This power–and the relative presence and absence of sources that result–carries through to the moment of archive (or dataset) creation. The point at which the French government decided which documents to preserve in their official records and which documents to discard further consolidated the colonial perspective on the events that transpired and why. Returning to the archive even centuries later, as Trouillot explains, one is still subject to those originary forces of power and more–a researcher can decide to tell a story one way or another, but when confronted with the one-sidedness of the colonial archive–the intentionality of its presences and absences–it becomes far more difficult to recover accounts outside of the colonial view. 

A similar (if less overtly colonial) process takes place in the process of dataset creation. One might think not only of datasets that derive directly from digitized colonial archives–the BNF, for example–but also of datasets like the Colossal Clean Common Crawl corpus, or C4, which explicitly documents its decision to include certain ``high quality'' web documents and exclude others. There is the immediate concern about which world language are captured through a dataset that derives from the Internet, which remains predominantly English language with users centered in the Global North. Beyond that, Gururangan, et al. \cite{gururangan2022whose} have shown how the quality filters imposed on the original Common Crawl dataset exhibit a preference for language associated with wealthier, more educated, urban ZIP codes, on top of the global Anglophone bias of the original dataset. This is the power in the making of sources and archives as well. 

As for the making of narratives and the making of retrospective significance, these layers of power compound. It becomes more difficult to pull out data that represents non-dominant groups, just as it requires yet more effort to develop narratives that reflect them. Trouillot’s atomization of the moments by which power enters the archive thus becomes a lens through which to refract the power that similarly overdetermines the construction of datasets, better attuning us to the contents of the archive and what stories they enable, and to the absences of the archive as well. These absences, in turn, point back to questions of power, fostering alertness to what stories the archive makes possible, and what stories the archive will forever foreclose.

This, too, is a lesson that might be mapped onto the data. Just as silences enter the dataset at the moment of its construction, so too do second-order silences enter at the moment of the making of narratives. Consider, as Bender et al. \cite{bender2021dangers} do in ``Stochastic Parrots,'' how the filtering of text relating to queerness and queer people in the C4 dataset results in a model lacking examples of queer language to draw upon when users (or researchers) prompt a large language model like GPT4 to generate liberatory examples of queer life; or how even datasets that are constructed with care and intention, such as the Colored Conventions Corpus \cite{cc-corpus}, which documents the meeting minutes of the aforementioned Colored Conventions–Black organizing meetings that took place during the nineteenth century, cannot be used to recover the voices or stories of the women organizers involved, since their contributions were not documented in their own time. Once again, we are confronted with what in archival theory is described as archival silence–the absences in the archive, both unintentional and by design–that limit what the archive can disclose.  

\subsection{Absence as Presence}
A second set of theories of the archive take up this issue of archival silence, instructing us on how we might reconceive the absences of the archive as commanding presence in and of themselves. While they cannot tell us what these figures said in their everyday conversations, where they went in their everyday activities, or how they truly lived their everyday lives, the absence of this information–and our own knowledge that it should exist–can hold open the space for what we do not and, at times, cannot ever know \cite{klein2013image}. Two concepts from the field of critical data studies are helpful here; the first is the idea of the null value, as articulated by Jacob Gaboury \cite{gaboury2018becoming} and extended by Jessica Marie Johnson \cite{johnson2018markup}. Drawing inspiration from the idea of how as-yet-filled database records are initialized with a null value, allocating memory 
for a value which may or may not be assigned in the future, Gabory theorizes structural queerness as a conceptual null value–a position that claims space without needing to be defined as a 0 or 1. Johnson takes Gaboury’s idea and extends it to the archive, figuring the empty records in the ledgers of the enslaved
as similarly space-holding. These empty records, often having to do with women and girls, in their own time and in the present demand that their lives be recognized, even as they lack identifying details. 

Here, Johnson also draws from the work of Saidiya Hartman \cite{hartman2008lose}, whose importance in this discussion is impossible to overstate. Hartman’s early work on the archive surfaces the compounding effects of what she calls the ``double bind'' \cite{hartman2022scenes} the ostensibly emancipatory efforts and corrections that ultimately reanimate the exercise of power against Black people and reinforce the conditions of chattel slavery \cite{hartman2008venus}. Hartman argues that one aspect of this accumulation of suffering stems from the retelling of those scenes of subjection, which never exceed their violent effects. She asks: ``How does one represent the various modes of practice without reducing them to conditions of domination or romanticizing them as pure forces of resistance?'' \cite{hartman2008venus} It is in this challenge to retell and represent without rehearsing pain that she grapples with historical absence. Even as the revival of stories that escape documentation remains an impossibility, the absence of those stories plays an active part in the writing of history (or ``counter-history'' \cite{hartman2008venus}); it becomes a presence.

This conditioning of an absent presence shares a cadence with José Muñoz's \cite{munoz2019cruising} concept of queerness, which he describes as an orientation to absence. In his now seminal text, \textit{Queering Utopia: The Then and There of Queer Futurity}, he states that queerness is ``a structuring and educated mode of desiring that allows us to see and feel beyond the quagmire of the present… Queerness is that thing that lets us feel that this world is not enough, that indeed something is missing'' (\cite{munoz2019cruising}, p.1). This ‘something’ that is missing marks absence as an affective site, a site for desire, and the beginnings of a path towards knowing that makes use of the ephemeral and the fleeting as meaningful markers of history that evade the devices of neat capture. For Muñoz, ephemera counts as, ``all of those things that remain after a performance, a kind of evidence of what has transpired but certainly not the thing itself'' (ibid, p.10). Perhaps then absence is able to designate desire. 

Desire of course is not unfamiliar to those working with information be it in archives or datasets, though it is seldom remarked upon. In the realm of information theory from Claude Shannon and Warren Weaver’s work to present, absence has been equated with that which is unpatterned, entropic, or without meaning culling forth desires for clarity, legibility, and completion. While absence reflects our desires back to us it also makes clear the normative structuring and residual legacies which come to define and organize not only what types of information are meaningful but what counts as information to begin with. 

Conversely, Muñoz’s embracing of desire as its own form of knowing might suggest a different response to absence, not as a site to be filled but with a reverence for other forms of sharing information beyond the index. It is a tacit acknowledgement of the limitation of the index to account for other ways of knowing. Those that can only exist on the horizon, not an arrival at a destination, but as attuned yearning suturing together contingent acts. For Muñoz, the performance, the dance floor, the exhaust of bodies in motion cannot be reconstructed through the materials left in their wake. Instead, the event itself is the moment by which information is shared through the repertoire of shared acts taken together. His insistence on the importance of ephemera as evidence of the event is insightful in that it marks a strong distinction between the event and the translation of the event into information. Absence becomes uniquely situated in that it makes our desires explicit and unavoidable. 

\subsection{Absence as Productive}
Lastly, we consider how the shift from bias to absence offers productive opportunities for the design of algorithmic systems. By labeling this move ``productive,'' we do not imply that the other two are not; rather, we emphasize the creativity and imagination this reading brings to our analytic work. With the term ``productive,'' we stay with the difficulty of speaking to nearby absence, working with absence not as a qualifier of truth or fact but instead as a provocation to cull forth a different halting problem. This analysis involves engaging in the slower work of caring for an absence, inviting reference.  
 
We begin by looking to Saidiya Hartman’s development of critical fabulation as an archival method that attempts to ``jeopardize the status of the event, to displace the received or authorized account, and to imagine what might have happened or might have been said or might have been done'' (\cite{hartman2008venus}, p.11) as a method that engages absence in a generative way. Hartman describes the nature of the process as ``a history written with and against the archive'' (\cite{hartman2008venus}, p.12). This entanglement with the archive reveals ways in which the power dynamics that are embedded in the archive, as Trouillot \cite{trouillot1995silences} describes, can be troubled by settling into the presence of its absences as places to question and imagine things differently. More specifically, while critical fabulation is structured through archival research, it is also speculative. This speculation is based on careful research and engagement with the fragments that have been captured in the archives, yet it attempts to push against the limits of them and the narratives that have been constructed from them. By occupying absence as an opportunity to illuminate both history and our present, narration offers an imagining that is critical, personal, and situated. It tells stories of real people, using their words when they are available.
For Hartman, for example, the two girls in ``Venus in Two Acts'' \cite{hartman2008venus}, or the young Black women at the turn of the twentieth century in Wayward Lives, Beautiful Experiments. Writing counter histories that are also speculative allows us to imagine within the absence while also accepting the ``incompleteness'' of the endeavor and ``the ongoing, unfinished and provisional character of this effort'' (\cite{hartman2022scenes}, p.14). As a result, we develop meaning beyond the limits of that which has been captured in the archive while maintaining the absent presence as a null value that cannot be ``solved,'' or ``corrected'' fully. 
 
Tiya Miles  \cite{miles2022all} deployment of critical fabulation 
in her research of Ashley’s sack offers possibilities for struggling with this absence and incompleteness. To begin, in 1921 Ruth Middleton embroidered a cotton sack with the story of her great grandmother Rose who gave the sack to her daughter Ashley, Ruth’s grandmother, when she was sold away from the plantation at age 9. If we consider Ashley’s sack with Trouillot’s moments there are two important points that highlight the productive potential of absence. First, Ruth’s embroidery on the sack is a moment of fact creation. It is a personal act that marks and captures the sack as a material connection between Rose, Ashley, and Ruth and the broader experience of Black women of their respective times. If Ruth doesn’t embroider, would we know about the sack? Would we understand its relationship between this family of women and the realities and aftermath of slavery? Ruth’s embroidery makes the sack identifiable to archival systems, leading to its addition to the exhibitions at Middleton Place and the Smithsonian National Museum of African American History \& Culture (NMAAHC) in Washington, DC. But it also makes other ``emergency packs'' and ``story cloths'' more visible in the broader stories of Black women that are missing or undertold within the archives. Ten lines of embroidered text offer a glimpse into the lives of the three women: Rose, Ashley and Ruth. Along with an inventory of the items that Rose filled the bag with, Ruth’s action to embroider their story on the sack captures not only the personal experiences of this family but also introduces a record of practices of love and care that connect Black women of a particular time and place \cite{miles2022all}. 

After seeing the sack at NMAAHC in 2016, Miles is prompted toward a``deeply exploratory and experimental project'' that attempts to learn as much as possible about the sack and its owners, culminating in the book \textit{All that She Carried: The Journey of Ashley’s Sack, a Black Family Keepsake}. While the exhibitions acknowledge Ashley’s sack as a source, Rose, Ashley, and Ruth remained partial. Miles’ explorations of the lives of the three women works to imagine within the absence of this partiality. 

Miles attempts to trace the lives of Rose, Ashley, and Ruth through Ashley’s sack not only as an artifact, but also as an archive by placing it ``in conversation with other sources and considering its various historical contexts'' (\cite{miles2022all}, p.17) in order to speak to a larger collective experience. Miles adapts both Fuentes’ archival strategy of reading ``along the bias grain'' and Hartman’s writing practice of critical fabulation as methods for searching for the experiences of enslaved Black women that push against the ``violence and distortion of traditional archives'' and archives that ``do not faithfully reveal or honor the enslaved'' \cite{miles2022all}. 
These conditions of absence are not confronted through correction, instead the orientation is a repositioning---a shift to the location of the women of focus, Rose, Ashley, Ruth, and other Black women engaging their collective experience, including Miles. Miles is sometimes narrator, sometimes researcher, sometimes subject, and even sometimes a daughter and grand-daughter. These moments of assembly extend Hartman’s critical fabulation as a method for imagining what could have been (and what could be). Consistent with Hartman’s formulation, the act of writing All that She Carried… is not giving voice to Rose, Ashley, and Ruth but it occupies the space of their absence in the archives – highlighting its presence and shifting our focus to it by anchoring to the situated story of these three women as part of a broader archival story.

Miles describes the story of Ashley’s sack as ``a quiet story of transformative love lived and told by ordinary African American women---Rose, Ashley, and Ruth...'' (\cite{miles2022all}, p.3). This story offers an important addition to the traditional archive. The woven sack, filled with its contents gives insight into the intersection of love and survival that Rose, Ashley, Ruth and other enslaved women and their descendants lived (and live) through their actions, words, and the labor of their bodies. The treatment of this personal artifact and story extends a Black Feminist practice that explores the connections between the personal and the collective. By telling these personal stories of individual Black women we can move beyond just cataloging their names or understanding them as data that bounds them as property or victims of violence.  By connecting Hartman’s writing practice of critical fabulation with Fuentes’ archival practice of reading along the bias grain, and Trouillot’s call to seek out the actual material things that “enslaved people touched, made, used, and carried—in order to understand the past,” Miles’ method reworks our understanding of what can be archival and offers examples of the possibilities in imagining within the absence. 
 

\section{Absence in/as Design Speculation}
Next we turn to four design speculations that each materialize and complicate one or more of the above three dimensions of absence (power, presence, and productive). By design speculations we refer to a specific tradition of speculating into what might have happened in the past in ways that draw a link to redressing bias in the present and the future \cite{hartman2008venus,sherman_black_2023,prado2018technoecologies}. Within this context, speculation works as exploratory and experimental---situated in, and occuring in relation to, the specific archive or dataset it engages. This distinguishes our approach from other speculative methods such as speculative design \cite{dunne2013speculative} and design fiction \cite{bleecker2022design} that tend to use design as a futuring activity and position a designer- or author- as the driver of that activity. By engaging existing works as speculative artifacts, we take up an interpretive stance, examining the work as a means of troubling the authority of the archival and data-driven systems they engage.

We organize these projects across two themes—one focused on AI-based language translation and the other on quilts—to examine two distinctly different relationships to algorithmic design. With each thematic pairing, we put imaginative reworkings of algorithmic bias in dialogue with critical archival concepts to examine what design scholarship might learn from their interrelation. 

While AI translation directly speaks to datasets (and data settings \cite{loukissas2019all}), quilts are less often approached as computational artifacts composed of data. By turning to both artifacts, we invite a more expansive reading of the dataset and re-examine the material and compositional boundaries of algorithmic design. 

\subsection{AI Language Translation}
\subsubsection{Design Speculation 1: Lacey Jacoby's AI Translation}
Our first example of translation work comes from design researcher Lacey Jacoby \cite{benabdallah2022slanted}, who reimagined Google Translate to reveal linguistic cultural bias often hidden within the tool. ``Welcome to a new translation experience'' the app announces. A drop down menu offers three sample phrases for translation from English to Japanese. Selecting the phrase ``What time is it?'' prompts the system to display the Japanese translation 
\begin{CJK}{UTF8}{min}今何時ですか？\end{CJK}
with a conjoining pop up box that asks ``Are you a 7-year old girl asking your mother what time it is?'' This question surfaces an assumption (or bias) built into Google Translate’s algorithm that makes decisions about the gender and seniority of the person speaking and the person spoken to. Rather than change the answer or algorithm that produced it, the designer made the hidden interpretation visible. 

With this experiment, we see Jacoby expose not only gendered and ageist assumptions within the algorithmic model, but also aspects of the dataset that do not exist. Based on its answer, the tool likely did not train its algorithmic model on data with an older man asking a young girl for the time. The absence of this data activates the acknowledgment of its loss. From the perspective of the designer, the data absence works as power, revealing a patriarchal and ageist tendency to assume young girls ask older men for facts. But from the perspective of a user, the absence also works as presence, pushing back against correction, which obfuscates the problem. 

\subsubsection{Design Speculation 2: Bettina Judd's Glossolalia}
We next put Jacoby’s provocation in conversation with the poetic experimentation of interdisciplinary writer, artist, and performer Bettina Judd and specifically her writing on ``Glossolalia''  \cite{judd2019lucille}, an engagement with poet Lucille Clifton’s process of automatic or spirit writing, a form of linguistic composition without conscious intention. In attempting to understand Clifton’s poetry, Judd finds it difficult to process the spirit writing—an exploration of, in her words, ``the space between knowing and unknowing'' (ibid, p.141). With glossolalia, or what she terms ``an experiment in speaking in tongues,'' she approaches Google Translate. Her process includes several phases: she speaks in tongues; transcribes her tongues; runs the transcript through Google Translate, which automatically detects Swahili; runs the output through the translation of multiple additional languages (in rough alphabetical order, e.g. Catalan, Basque, Azerbaijani); runs the translation back to English; and finally composes poetry based on those outputs. In Judd’s staged poetry performance, the absence of linguistic recognition by the translation algorithm poses the impossibility of knowing Clifton’s work. As a contemplation on knowing, the absence becomes productive. It supports Judd in creating a response to Clifton in the form of poetry.  

\subsection{Quilt}
\subsubsection{Design Speculation 3: Encoding in the Freedom Quilts}
Our third design speculation focuses on experimentation with quilts and specifically Freedom Quilts, the covert system of encodings that charted escape routes for enslaved Black people.
Elsewhere we have described the Freedom Quilts as ``a vital form of computation that forces us to rethink what computing can be when freed from its dependence on colonial pursuits of managing bodies, spaces, and resources'' \cite{morrison2022voluptuous}. Though we embrace the Freedom Quilts as a Black and diasporic form of computing, they also live in tension, burdened by taxonomic logics that demand evidence and proof. This contested history of the Freedom Quilts challenges the distinctions between absence and legibility, turning our attention to the use of protocols that cannot be transparent but must circulate through opacity and the embedded trust of interdependence. The absence of definitive proof of the quilting patterns, the protocols for their assemblage, the exact location and indexing of use, these details distract from the provocation that they offer our current inquiry on absence. Instead, the Freedom Quilts thrive as a clandestine network of information sharing because they are rendered absent within the imagination of the plantation system, obsessed with endless pursuits of measurement, hierarchy, logistics, and analytics \cite{rosenthal2019accounting}. If its specific patterns and symbols guided those escaping slavery to safe houses or offered instructions and warnings about the journey on the Underground Railroad, they also help us think about the encoding of data and the protocols of trust building. This tarries the lines between no longer useful distinctions between what is deemed computational and what is social and relational. The acts of computing in this context, the calculating of trajectories and distance, the encoding of geographic information, the use of pattern recognition, are not separable from the social relations that undergird their function. Instead they are contingent upon recurring acts of interpretation, negotiation, trust building, and encoding.  

As such the nature of code is being written through the creolization of symbols, meanings, context, and codes that change and shift in accordance with different sites and ensembles. This complicates our familiarity with programming as the temporality of code, as executable function, as cause and effect. Because transparency comes with a price for some, the clarity of code as executable function---able to be read and performed void of context, was untenable. Instead, the shifting nature of the code required a great deal of maintenance which further emboldened and fortified the social infrastructures on which the quilts ran. While encountering quilts left in public fugitives would discern the code and simultaneously have to read it in context, within the geography of placement. In this instance the executability of code is halted as a declarative axiomatic language imagined within syntax. Code is not an absolute instruction but is contingent and contextual read in addition to landscape and relationships.

\subsubsection{Design Speculation 4: Curry Hackett's Reimagining of the Gee's Bend Quilters}
The non-totalizing concept of computation from the Freedom Quilts above opens important lines of interpretation for the work of architect Curry Hackett. As our fourth design speculation, we turn to his series of generative AI quilted pattern renderings, inspired by the quilts from Gee’s Bend, the set of intricately assembled asymmetrical improvisational patterns designed by Black artisans from a small Alabama town, Hackett conjures (impossible) cities. 

In one of Hackett's images we see two Gee’s Bend quilters, one wearing a floral print dress and the other in a loose fitting yellow jumpsuit, standing on a platform suspended above a skyscraper in a dense urban setting. But instead of using the scaffolding to clean windows, apply paint, or repair exterior decay, the women appear to be installing one of their quilts. A colorful patchwork made of interspersed rectangular scraps, some patterned and others solid, covers the visible length of the city high rise. The woman in the floral dress looks as if she’s about to climb higher, swinging herself onto the next set of scaffolding to continue her installation. 

Using the quilt to rethink architectural space, Hackett’s imagery complicates what counts as ordered and spatially secure. Seeing a soft tapestry atop a towering skyscraper feels both exciting and off kilter. Compared with a stereotypical generative AI output for a ``city of the future'' prompt, the minimalist landscape awash with technological chrome and white pedestrians, Hackett’s image accomplishes a re-centering of Black cultural production that he describes above. It elevates the Gee’s Bend quilters both literally and figuratively, recognizing their pattern work as part of a missing dataset in the visualization of urban spatial innovation. He explains of the quilts:

\begin{quote}
``I see this work as an open source repository for free ideas for how to imagine a world that centers Blackness. You know, Black culture, Black aesthetics, Black modes of living.'' \cite{hackett-02}
\end{quote}

But like the Freedom Quilts, the quilted skyscraper does more than expose the absence as power or even presence. It also helps broaden the normative aesthetics of pattern. Like the maintenance of the Freedom Quilts, the Gee's Bend aesthetics in Hackett’s renderings evoke a powerful call to improvisation and contingency. In their aesthetic ordering, contingency---the interdependence of elements---becomes the recursive practice for computing. This practice prefigures contemporary contingent turns in computer science and media studies \cite{fazi2018contingent,parisi2022recursive}. It offers a challenge to a narrow and often dangerous view of pattern as a systematic ordering that makes architecture legible and useful. Here we find a more expansive and expressive reading of architectural order that approaches the absence as productive. The quilted skyscraper, in line with the Freedom Quilt, is an object whose troubled existence opens the possibility for a radical reimagining of computational bias in the form of possibility. 

\section{Discussion}
If we are to pursue absence as a meaningful concept to extend beyond the limitations of bias, we must wrestle with its commonplace reading as the negative space needing to be filled or corrected. Instead, we wonder what absence might provoke in our instrumental approaches to information that halt desires for a total or completed set and emphasize our considerations for building relationships to the object or dataset. Put simply, what is to be done with absence? How can our desires for what is not present reorient our ability to work with data? How then, might the historical, the then and there, proffer a different, here and now? 

Across historiography, gender studies, Black studies, and queer theory, absence registers differently, not always as a space for the unknown, but also as a vital signification towards knowing. What appears mute, insists on this enhanced attunement. The absence is not a silence, not a missing or non-existent sensing. Rather it establishes contact and connection with an analyst at a level that requires effortful engagement.  

If absence works as a site for fabulation (Hartman), desire (Muñoz), and elasticity (Fuentes), then absence may be able to forge a different relationship to evidence, not as the inverse of presence, but as a way to mark the space for what is to come. The aspirations for producing knowledge from the archive or the dataset often hinge on a sense of completion, to equate volume with truth. We solve for what we can show, and the proof is self-evident. In these equivalences is a latent voice that speaks of a desire to know. So if desire is not absent from the work that the archive or dataset produces, what is to be done with it? 

From the above design speculations we see a concern for how power, presence, and productive(ness) work as de facto sites of meaning making. The translation app and the quilt both offer spaces for rereading dataset silences as invitations for alternative modes of engagement: intimate listening to the contours of entrenched gender norms; attunement to formats for decoding and realigning pattern; resurrection of connections to the inner self. It is inside the null value that we find new and different value by allowing something unexpected and generous to escape. 

Like the experiments in AI translation, the Gee's Bend and Freedom quilts perform a particular instantiation of diasporic code and the social relations of computing. Yet they are not exceptional; they continue a longer lineage of quilts as informational interfaces that store and transmit data. Bill Arnett in his text, ``Souls Grown Deep,'' writes that ``Quilts represent one of the most highly evolved systems of writing in the New World. Every combination of colors, every juxtaposition or intersection of line and form, every pattern, traditional or idiosyncratic, contain data that can be imparted in some form or another to anyone. All across Africa, geometric designs, the syntax of quilt tops, have been used to encode symbolic or secret knowledge'' \cite{tobin2000hidden}. Quilts, the assembling of patterns to store and encode information act as a refraction, altering and resituating the centrality of pattern in the words of Claude Shannon, Warren Weaver, and Norbert Wiener. As noted earlier, pattern emerges from within postwar information theory as the primary metric for structuring and transmitting information. However, the aesthetics of what constitutes a pattern were largely left absent. In this absence pattern was discussed as singular and universal, deploying a preference for legibility, uniformity, and repeatability. This is what separated signal from noise, ensuring that the aesthetics of pattern remained absent. The above design speculations, and particularly the Freedom Quilts, call this assumed and normative aesthetic ordering into question. For aesthetics are not simply a question of style or appearance but indicative of the foundational protocols that define and determine taxonomies of order \cite{wynterrethinking}. In part this aesthetic order determines what counts as information and what falls beyond the bounds, conceivable only as noise or absence. As we argue here, absence is not the lack of presence but can also hold the noisy spaces for possibility resounding with the din of improvised relations. 

We also find that with the shift toward absence, the design speculations operated between the traces and fragments that have been captured---against the grain and inside the null value, within the absence presence. They help us, as analysts of design, consider what might have happened where the record is incomplete, compromised, or inaccurate. But they also challenge a linear reading of speculative time, one narrowly focused on the past conceived as a point that has already occurred and ended. To do this reorientation, they frame the past as entangled with present realities and future possibilities. They reveal an intermingling that troubles distinctions of history and future. As a result, the speculations are as A.P. Gumbs describes \textit{speculative documentary}---both ancestrally cowritten and collaboratively rendered with survivors as ``far-into-the-future witnesses to the realities we are making possible or impossible’’ (\cite{gumbs2020m}, p.xi).

It is at the seams of this reimagination that we look to absence as ab-sense. A breaking apart of absence into ab-sence foregrounds the prefix ``ab'' (meaning away or from) alongside the ``sense'' of embodied perception. To work from embodied perception is to recognize the always implicated and implicating stance of the viewer/analyst and the need to reckon with objectivity and forced distance with data. In this intimate relation, absence invites an attunement to bodies in motion, to encounters that begin from the sensory, and a particular bodily knowing. It reveals experiences encoded in archives, databases or technologies as always tied to lives marked by ideology.

It’s not without ambivalence that we present this shift in thinking. A move toward absence matters for conversations on data bias but it is not possible to solve this work or resolve the harm that has taken generations to build and is premised on trauma. Neither can this theoretical tradition be ‘imported’ from critical archival theory into data practices. These reorientations in method are not design guidelines for citational practices, nor can they be engaged quickly or wholly. They ask us, as data analysts, to care for the absence of something without determining it to be what we are projecting onto it. They urge us to cultivate ways of staying within the limitations of what the absence is trying to do. By nurturing an engagement with absence within data practices, we recognize data as just the surface, as the stuff/symptom concealing the people and relationships behind them. Absence helps us probe our methods for acknowledging underlying and longstanding forces of capital and colonial power by reimagining their role within and beyond the worlds they try to contain. 

\section{Conclusion}
Our aim with this essay has been to examine the shift from algorithmic bias to algorithmic absence as a means of enriching ongoing analysis of data practices and addressing their systemic effects. Scholarly critiques of bias have exposed the imperative to solve systemic problems of algorithmic harm or exclusion by filling a gap. This gap-filling presents a bolstering of big-data fantasies based on the capturing of more data (in volume, diversity, or otherwise).

Across our investigations of absence, we see how the concept becomes: an indicator of power with which analysts can carefully work through what is not there; an activation of presence that allows analysts to wrestle with what gets hidden, excluded, or left incomplete as a lively contributor to meaning making; and a capacious idea of the productive, which opens analysts to imaging the world differently. Together these dimensions of absence prompt us, as analysts, to consider the nature of the agreements and conditions we need to do our work with intention, accountability, and care. They draw attention to the social relationships around knowledge produced through and with algorithmic design.

\section{Acknowledgements}
We thank our colleagues in the UW TAT Lab and UW Simpson Center for the Humanities ``Reimagining Datafication'' research cluster as well as funding support offered by the UW Simpson Center and NSF grants 2222242, 2310515, and 2210497. We also thank Afroditi Psarra, who co-caught the \textit{On the Bias} course that led us to Lacey Jacoby's AI translation project and discussions with Chari Glogovac-Smith, Gabrielle Benabdallah, and Bettina Judd.



\bibliographystyle{ACM-Reference-Format}
\bibliography{refs}

\appendix









\end{document}